\documentclass[prb,showpacs,superscriptaddress,floatfix,twocolumn]{revtex4}
\usepackage{amsfonts}
\usepackage{stmaryrd}
\usepackage{bbm}
\usepackage{mathrsfs}
\usepackage{tipa}
\usepackage{amssymb}
\usepackage{txfonts}
\usepackage{graphicx}
\usepackage{dcolumn}
\usepackage{epstopdf}
\usepackage{array}

\newcommand{\PreserveBackslash}[1]{\let\temp=\\#1\let\\=\temp}
\newcolumntype{C}[1]{>{\PreserveBackslash\centering}p{#1}}
\newcolumntype{R}[1]{>{\PreserveBackslash\raggedleft}p{#1}}
\newcolumntype{L}[1]{>{\PreserveBackslash\raggedright}p{#1}}

\begin{document}

\newcommand*{\cm}{cm$^{-1}$\,}

\title{Evidence of topological insulator state in the semimetal LaBi}

\author{R. Lou}
\affiliation{Department of Physics and Beijing Key Laboratory of Opto-electronic Functional Materials $\textsl{\&}$ Micro-nano Devices, Renmin University of China, Beijing 100872, China}

\author{B.-B. Fu}
\author{Q. N. Xu}
\affiliation{Beijing National Laboratory for Condensed Matter Physics, and Institute of Physics, Chinese Academy of Sciences, Beijing 100190, China}

\author{P.-J. Guo}
\affiliation{Department of Physics and Beijing Key Laboratory of Opto-electronic Functional Materials $\textsl{\&}$ Micro-nano Devices, Renmin University of China, Beijing 100872, China}

\author{L.-Y. Kong}
\author{L.-K. Zeng}
\author{J.-Z. Ma}
\affiliation{Beijing National Laboratory for Condensed Matter Physics, and Institute of Physics, Chinese Academy of Sciences, Beijing 100190, China}

\author{P. Richard}
\affiliation{Beijing National Laboratory for Condensed Matter Physics, and Institute of Physics, Chinese Academy of Sciences, Beijing 100190, China}
\affiliation{School of Physical Sciences, University of Chinese Academy of Sciences, Beijing 100190, China}
\affiliation{Collaborative Innovation Center of Quantum Matter, Beijing, China}

\author{C. Fang}
\affiliation{Beijing National Laboratory for Condensed Matter Physics, and Institute of Physics, Chinese Academy of Sciences, Beijing 100190, China}

\author{Y.-B. Huang}
\affiliation{Shanghai Synchrotron Radiation Facility, Shanghai Institute of Applied Physics, Chinese Academy of Sciences, Shanghai 201204, China}

\author{S.-S. Sun}
\author{Q. Wang}
\affiliation{Department of Physics and Beijing Key Laboratory of Opto-electronic Functional Materials $\textsl{\&}$ Micro-nano Devices, Renmin University of China, Beijing 100872, China}

\author{L. Wang}
\author{Y.-G. Shi}
\affiliation{Beijing National Laboratory for Condensed Matter Physics, and Institute of Physics, Chinese Academy of Sciences, Beijing 100190, China}

\author{H. C. Lei}
\author{K. Liu}
\affiliation{Department of Physics and Beijing Key Laboratory of Opto-electronic Functional Materials $\textsl{\&}$ Micro-nano Devices, Renmin University of China, Beijing 100872, China}

\author{H. M. Weng}
\affiliation{Beijing National Laboratory for Condensed Matter Physics, and Institute of Physics, Chinese Academy of Sciences, Beijing 100190, China}
\affiliation{Collaborative Innovation Center of Quantum Matter, Beijing, China}

\author{T. Qian}
\affiliation{Beijing National Laboratory for Condensed Matter Physics, and Institute of Physics, Chinese Academy of Sciences, Beijing 100190, China}
\affiliation{Collaborative Innovation Center of Quantum Matter, Beijing, China}

\author{H. Ding}
\affiliation{Beijing National Laboratory for Condensed Matter Physics, and Institute of Physics, Chinese Academy of Sciences, Beijing 100190, China}
\affiliation{School of Physical Sciences, University of Chinese Academy of Sciences, Beijing 100190, China}
\affiliation{Collaborative Innovation Center of Quantum Matter, Beijing, China}

\author{S.-C. Wang}
\email{scw@ruc.edu.cn}
\affiliation{Department of Physics and Beijing Key Laboratory of Opto-electronic Functional Materials $\textsl{\&}$ Micro-nano Devices, Renmin University of China, Beijing 100872, China}

\begin{abstract}
  By employing angle-resolved photoemission spectroscopy combined with first-principles calculations, we performed a systematic
  investigation on the electronic structure of LaBi, which exhibits extremely large magnetoresistance (XMR), and is theoretically
  predicted to possess band anticrossing with nontrivial topological properties. Here, the observations of the Fermi-surface topology
  and band dispersions are similar to previous studies on LaSb [Phys. Rev. Lett. \textbf{117}, 127204 (2016)], a topologically trivial
  XMR semimetal, except the existence of a band inversion along the $\Gamma$-$X$ direction, with one massless and one gapped Dirac-like
  surface state at the $X$ and $\Gamma$ points, respectively. The odd number of massless Dirac cones suggests that LaBi is analogous
  to the time-reversal $Z_2$ nontrivial topological insulator. These findings open up a new series for exploring novel topological
  states and investigating their evolution from the perspective of topological phase transition within the family of rare-earth 
  monopnictides.
\end{abstract}


\maketitle

Exploring exotic topological states has sparked extensive research interest, both theoretically and experimentally, due to 
their promising potential in low consumption spintronics devices \cite{Hasan2010,Qi2011,Weng2014}. In the past decade, since 
the discovery of quantum spin Hall effect in graphene \cite{KaneGraphene}, remarkable achievements have been reached, including
the findings of two-dimensional (2D) \cite{Bernevig2006,Konig2007,Knez2011} and three-dimensional (3D) topological insulators
(TIs) \cite{Chen2009}, node-line semimetals \cite{Burkov2011,Yu2015}, topological crystalline insulators \cite{Fu2011,Heish2012NC},
and Dirac and Weyl semimetals \cite{Wang2012,Liu2014,Weng2015Weyl,Lv2015,Huang2015,Xu2015Weyl}. Strikingly, although the TIs have
ignited the whole field for years, the realization of novel massless surface Dirac fermions, $i.e.$, topological surface states
(SSs), is still pretty exciting due to the great prospects in tunability of the topological characteristics through easily
accessible manipulations \cite{Ando2013}. Among the extensive efforts, the study on the topological phase transition (TPT) can
also effectively facilitate the exploration and investigation of topological SSs and even the Dirac semimetals \cite{Sato2011,
Xu2011,Lou2015}.

\begin{figure}[!t]
  \setlength{\abovecaptionskip}{-0.45cm}
  \begin{center}
  \includegraphics[width=1\columnwidth]{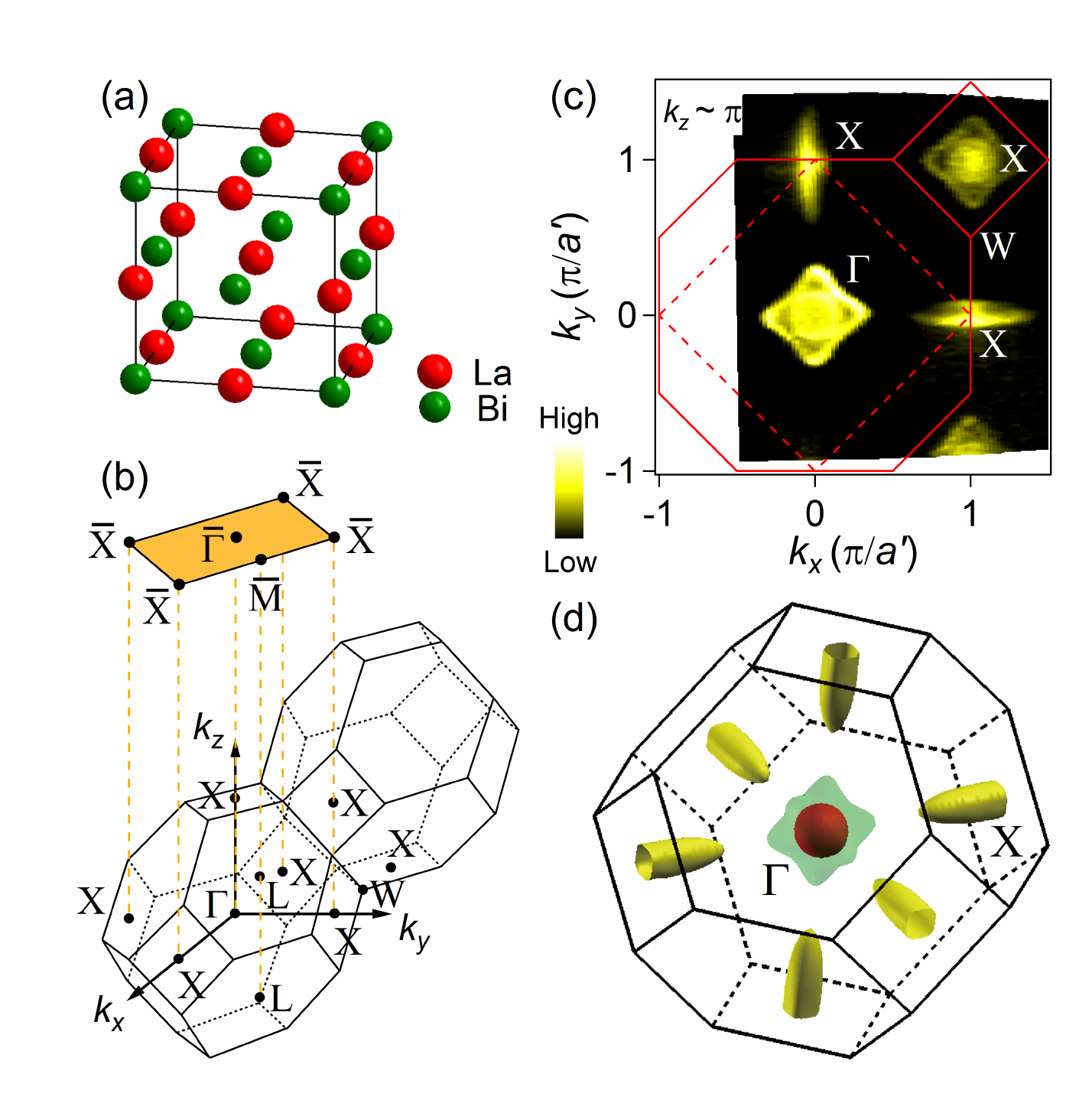}
  \end{center}
  \caption{(Color online) Crystal structure and FS topologies of LaBi.
  (a) Crystal structure of LaBi.
  (b) Bulk BZ and the (001)-projected surface BZ of LaBi.
  (c) FS intensity plot of LaBi recorded with $h\nu$ = 84 eV, corresponding to the $k_z$ $\sim$ $\pi$ plane, obtained by integrating the
      spectral weight within $\pm$10 meV with respect to $E_F$. $a'$ is the half of the lattice constant $a$ of the face-center-cubic unit
      cell. Red solid and dashed lines represent the 3D BZ and 2D surface BZ, respectively.
  (d) Calculated 3D FSs using the PBE functional.
  }
\end{figure}

\begin{figure*}[htbp]
  \setlength{\abovecaptionskip}{-0.5cm}
  \begin{center}
  \includegraphics[width=1.77\columnwidth]{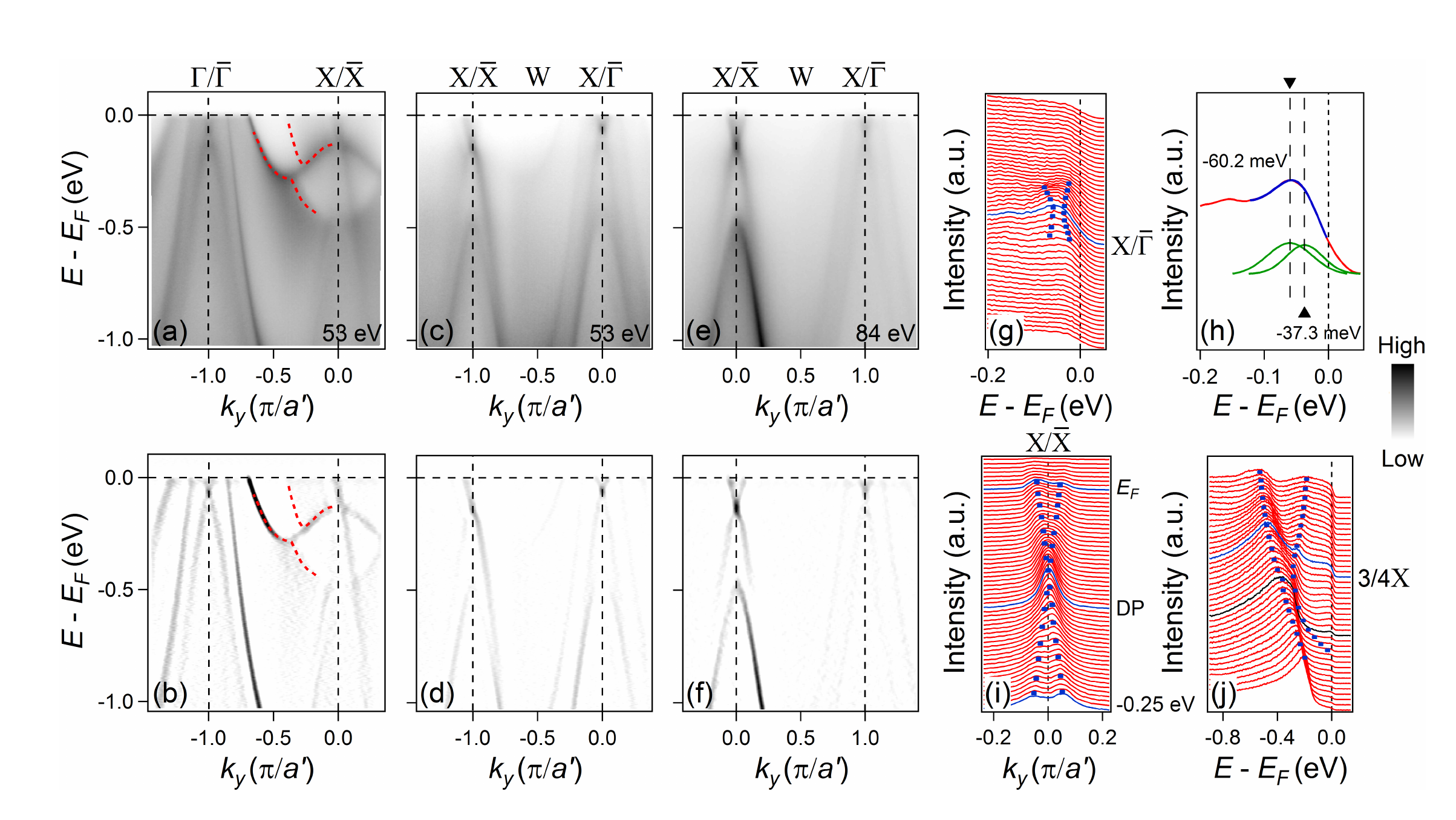}
  \end{center}
  \caption{(Color online) Band dispersions along high-symmetry lines taken at the $k_z$ $\sim$ 0 ($h\nu$ = 53 eV) and $k_z$ $\sim$ $\pi$
                          ($h\nu$ = 84 eV) planes of LaBi.
  (a),(b) Photoemission intensity plot along the $\Gamma$-$X$ direction and corresponding second derivative plot recorded with $h\nu$ =
          53 eV, respectively. Red dashed curves are guides to the eye for the band anti-crossing feature.
  (c),(d) Same as (a),(b) but along the $X$-$W$-$X$ directions.
  (e),(f) Same as (c),(d) but measured with $h\nu$ = 84 eV, displaying more prominent surface band features around the left $X$ point.
  (g) EDC plot of (e) around the right $X$ point.
  (h) Single EDC at the right $X$ point in (e), highlighted by blue curve in (g). By fitting the EDC with two Lorentzian profiles, shown
      as green curves, we extract the peak positions, marked by two vertical dashed lines, and determine the presence of a surface band
      gap of $\sim$23 meV. The blue curve is the superimposed fitting result.
  (i) MDC plot of (e) around the left $X$ point. The binding energy of DP is $\sim$-0.14 eV, as highlighted by the blue curve.
  (j) EDC plot of (a) to demonstrate the existence of band anti-crossing and bulk band gap. A shoulder on the outer hole band is resolved
      and highlighted by a black curve.
  Blue dots in (g), (i), and (j) are extracted peak positions, serving as guides to the eye.
  }
\end{figure*}

Recently, the discovery of simple rock salt rare-earth monopnictides Ln$X$ (Ln = La, Y, Nd, or Ce; $X$ = Sb/Bi) \cite{Tafti2015,Sun2016,
Kumar2016,Yu2016,Pavlosiuk2016,Wakeham2016,Alidoust2016} has renewed the platform for searching these novel topological states. The most
remarkable signature of this series is the extremely large magnetoresistance (XMR) with a resistivity plateau at low temperatures, which
is proposed as the consequence of breaking time-reversal symmetry \cite{Tafti2015}. However, the theoretical predictions for the topological
properties vary in the Ln$X$ series \cite{Zeng2015}. In addition, similar fingerprints have also been reported in several semimetals including
TmPn$_2$ (Tm = Ta/Nb, Pn = As/Sb) \cite{WangK2014,Shen2016,WuD2016,XuC2016,WangY2016,WangZ2016}, ZrSiS \cite{Singha2016,Ali2016,WangX2016},
WTe$_2$ \cite{Ali2014}, Cd$_3$As$_2$ \cite{FengJ2015,Liang2015}, TaAs \cite{HuangX2015}, and NbP \cite{Shekhar2015}. The topological origin
of the large magnetoresistance and resistivity plateau is still under debate. Furthermore, the rich and interesting topological phases
predicted in the Ln$X$ family offer an unprecedented opportunity for investigating the novel topological properties and TPT in a relatively
simple system without convoluted bulk and SSs \cite{Zeng2015,JiangJ2015,Pletikosic2014,Schoop2016,Lou2016ZST}. To our knowledge, although
there have been several angle-resolved photoemission spectroscopy (ARPES) studies on this series of compounds \cite{Alidoust2016,Niu2016,
WuY2016,Nayak2016,Neupane2016} stimulated by previous investigations on LaSb \cite{Lou2016LaSb}, the experimental data and understandings
are still ambiguous to fully determine the existence of nontrivial band topology in other compounds of the Ln$X$ family.

In this paper, we report systematic ARPES measurements and first-principles calculations on LaBi single crystals. We identify two hole-like
Fermi surfaces (FSs) at the Brillouin zone (BZ) center $\Gamma$ and one electron-like FS at the BZ boundary $X$, similar to that of LaSb.
Furthermore, we find that LaBi exhibits a clear nontrivial band anticrossing along the $\Gamma$-$X$ direction, with the presence of one
massless and one gapped Dirac-like SSs around $X$ and $\Gamma$, respectively. Based on the observation of band inversion and odd number of
massless Dirac cones, our results unambiguously demonstrate the $Z_2$-type nontrivial band topology of LaBi. The diverse topological phases
among the Ln$X$ series of compounds provide an excellent platform for exploring novel topological states and investigating their evolution
from the perspective of TPT.

Single crystals of LaBi were grown by the In flux method. The residual resistance ratio [RRR = $R$(300 K)/$R$(2 K) = 204] and large
magnetoresistance (MR = 3.8$\times$10$^4$ \% at 2 K under 14-T magnetic field) indicate the high quality of samples used in this paper
\cite{Sun2016}. The Vienna $\emph{ab initio}$ simulation package is employed for most of the first-principles calculations. The generalized 
gradient approximation of Perdew-Burke-Ernzerhof (PBE) type is used for the exchange-correlation potential \cite{Perdew1996}. Spin-orbit 
coupling is taken into account. The $k$-point grid in the self-consistent process is 11$\times$11$\times$11. To get the tight-binding 
(TB) model Hamiltonian, we use the package WANNIER90 to obtain maximally localized Wannier functions of $d$ and $f$ orbits of La and $p$ 
orbit of Bi. ARPES measurements were performed at the Dreamline beamline of the Shanghai Synchrotron Radiation Facility with a Scienta D80 
analyzer, at the Surface and Interface Spectroscopy beamline of Swiss Light Source using a Scienta R4000 analyzer, and at the beamline 13U 
of the National Synchrotron Radiation Laboratory equipped with a Scienta R4000 analyzer. The energy and angular resolutions were set to 15 
meV and 0.2$^{\circ}$, respectively. Fresh surfaces for ARPES measurements were obtained by cleaving the samples $\emph{in situ}$ along the 
(001) plane in a vacuum better than 5$\times$10$^{-11}$ Torr. All data shown in this work were recorded at $T$ = 30 K.

LaBi crystallizes in a NaCl-type crystal structure (face-center cubic) with space group $Fm$-3$m$, in which Bi is located at the face center
adjacent to La atoms, as illustrated in Fig. 1(a). The schematic bulk Brillouin zone (BZ) and the (001)-projected 2D surface BZ are presented
in Fig. 1(b). One bulk $X$ and two bulk $L$ points are projected to the surface $\bar{\Gamma}$ and $\bar{M}$ points, respectively; two bulk
$X$ points are projected to the surface $\bar{X}$ point. Figure 1(c) demonstrates the FS topologies of LaBi recorded with photon energy $h\nu$
= 84 eV, close to the $k_z$ = $\pi$ plane according to our photon energy dependent study discussed below. One can obtain remarkable consistency
between the experimental FSs and theoretical calculations with the PBE functional \cite{GuoP2016}, as shown in Fig. 1(d), including one elliptical
electron pocket at $X$ elongated along the $\Gamma$-$X$ direction, and two hole pockets centered at $\Gamma$ with the intersecting-elliptical one
enclosing the circular one. Further examining the measured FSs carefully, we can observe additional FSs around $X$ perpendicular to the elongated
pocket along the $\Gamma$-$X$ direction, which is a common feature in the Ln$X$ series of compounds; this could be possibly interpreted by the band
folding effect as a consequence of breaking translational symmetry from bulk to the (001) surface \cite{Lou2016LaSb}. Moreover, the $k_z$ broadening
effect caused by the short escape length of the photoelectrons excited by the vacuum ultraviolet light in our ARPES experiments may also be an
alternative explanation \cite{Kumigashira1997,Kumigashira1998,Strocov2003,Niu2016}.

\begin{figure}[!t]
  \setlength{\abovecaptionskip}{-0.5cm}
  \begin{center}
  \includegraphics[width=1\columnwidth]{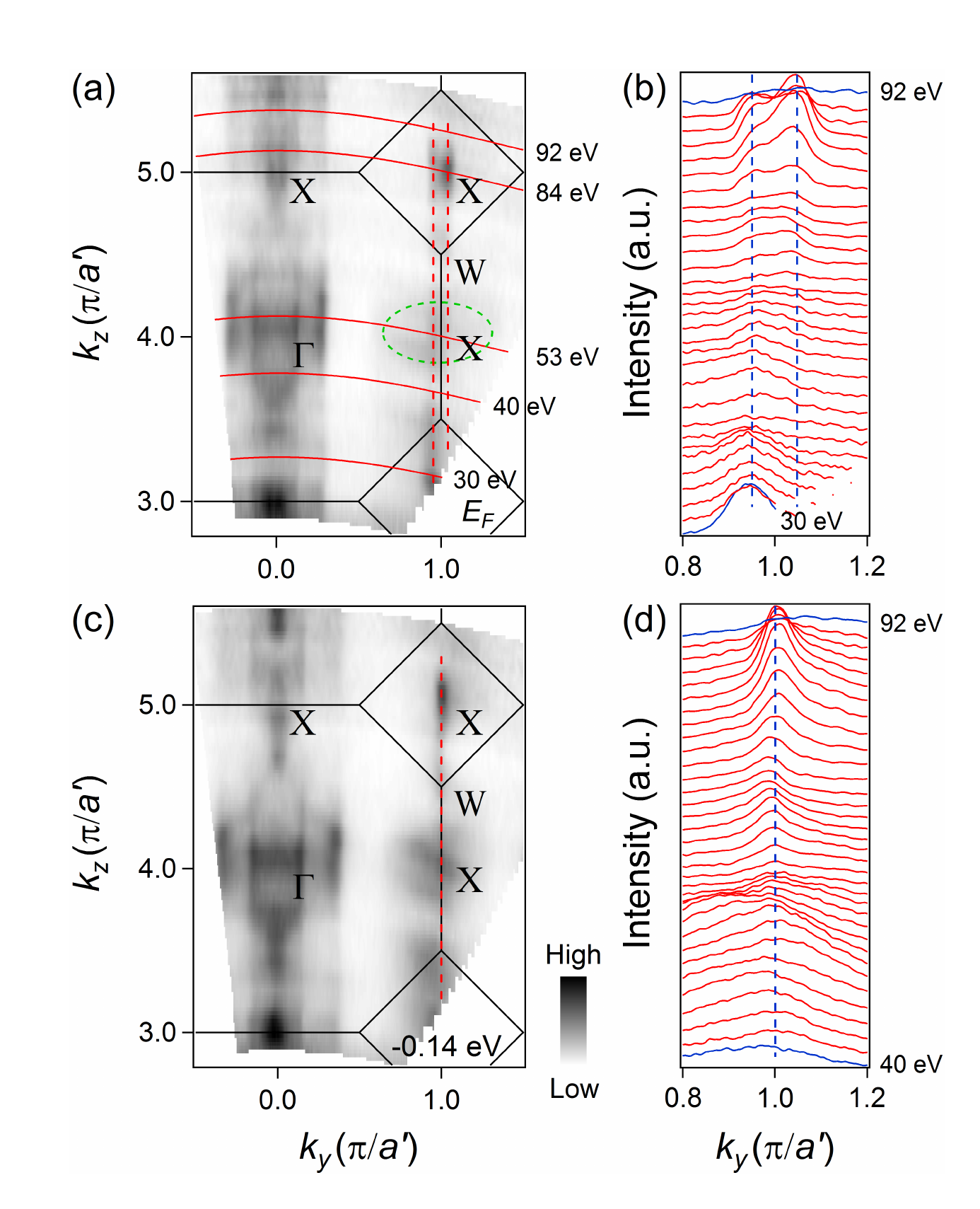}
  \end{center}
  \caption{(Color online) Photon energy dependence of the band structure along the $\Gamma$-$X$ direction of LaBi.
  (a) Photoemission intensity plot at $E_F$ in the $k_y$-$k_z$ plane at $k_x$ = 0 measured with photon energies from 20 to 100 eV.
      Red curves from bottom to top indicate the momentum locations taken at $h\nu$ = 30, 40, 53, 84, and 92 eV, respectively. The
      inner potential is estimated to be 14 eV. Green dashed ellipse represents the electron-like FS at $X$.
  (b) MDC plot of (a) around the BZ boundary.
  (c) Same as (a) but at $E$ = -0.14 eV, the binding energy of the DP at $X$.
  (d) MDC plot of (c) around the BZ boundary.
  Red and blue dashed lines in (a),(b) and (c),(d) demonstrate the $k_z$ independence of the Fermi crossings and DP of the SS around
  $X$, respectively.
  }
\end{figure}

In order to illuminate the underlying topological properties of the measured electronic structure of LaBi, we have investigated the
near-$E_F$ band dispersions along the high-symmetry lines $\Gamma$-$X$ and $X$-$W$-$X$, with photon energies $h\nu$ = 53 and 84 eV,
close to the $k_z$ = 0 and $\pi$ planes, respectively. Figures 2(a) and 2(b) show the measured band structure and corresponding second
derivative intensity plot along the $\Gamma$-$X$ direction recorded with $h\nu$ = 53 eV, respectively. The overall bulk band dispersions
are similar to our previous study on LaSb \cite{Lou2016LaSb}, except that the outer hole band around $\Gamma$ exhibits a shoulder at
$\sim$-0.33 eV when gradually leveling off from $\Gamma$ to $X$. Additionally, unlike the full parabolic electron band centered at $X$
in LaSb, the one in LaBi curves upward towards $X$ around the shoulder, forming a hole band with a top at $\sim$-0.14 eV at $X$. The
calculated band structure using the PBE functional can reproduce these bulk features very well and indicate that LaBi is a topologically
nontrivial material with band anticrossing along the $\Gamma$-$X$ direction \cite{GuoP2016}. Using the detailed energy distribution
curve (EDC) analysis presented in Fig. 2(j), we clearly observe a band gap around the shoulder, further demonstrating the existence of
nontrivial band inversion in LaBi. More ARPES spectra showing much clearer features around the shoulder and the electron band centered
at $X$ measured on another piece of sample can be found in the Supplemental Material \cite{Supplementary}.

Besides the expected bulk bands, there are some additional Dirac-like surface bands around the $\Gamma$ and $X$ points. To avoid
the complexity introduced by the convoluted bulk states, we investigate the electronic structure along the $X$-$W$-$X$ direction
with $h\nu$ = 53 and 84 eV in Figs. 2(c)-2(i). Owing to the projection over a wide range of $k_z$ \cite{Niu2016}, the Dirac-like
surface band around the bulk $\Gamma$ point in the $k_z$ = 0 ($\pi$) plane is projected to the bulk $X$ point in the $k_z$ = $\pi$
(0) plane, further validated by the surrounding two dispersive hole bands around the right $X$ points in Figs. 2(c)-2(f), which
are the contributions from the bulk $\Gamma$ point. Thus the surface bands located at $\Gamma$ points in Figs. 2(a) and 2(b) are
identical to those around the right $X$ points in Figs. 2(c)-2(f). Accordingly, it seems that the total number of Dirac points
(DPs) below $E_F$ is even, $i.e.$, 2, which is inconsistent with the odd number of band anticrossings in LaBi.

To examine this topological characteristic, we perform EDC and momentum distribution curve (MDC) studies around the right and left $X$
points in Fig. 2(e), and present the results in Figs. 2(g) and 2(i), respectively. Unlike the massless Dirac-like surface band with a
DP at $\sim$-0.14 eV in Fig. 2(i) (see detailed EDC plot in the Supplemental Material \cite{Supplementary}), which coincides with the
valance-band maximum at the $X$ points in Figs. 2(a) and 2(b), we clearly resolve an energy gap in Fig. 2(g) separating the upper and
lower linearly dispersive bands. Correspondingly, as seen in Fig. 2(h), the EDC at the right $X$ point in Fig. 2(e) exhibits a shoulder
beside the prominent peak. We use two Lorentzian curves to fit this single EDC and the fitting result is shown as the superimposed blue
curve. By extracting the peak positions of Lorentzian curves, we can obtain a surface band gap of $\sim$23 meV, suggesting the existence
of a massive Dirac cone at the bulk $\Gamma$ point. This exotic mass acquisition of Dirac fermions in LaBi was also observed recently by
Wu $\emph{et al}$. \cite{WuY2016}, demonstrating the novelty and complexity of the topological properties in this compound. Since the SS
at $\Gamma$ is proved to be gapped, the total number of massless Dirac cones below $E_F$ is odd. Therefore, by summarizing the bulk and
surface band dispersions along the high-symmetry lines, we conclude that LaBi is analogous to the time-reversal $Z_2$ nontrivial TI
\cite{Moore2007,Fu2007}, while the massive Dirac fermion at $\Gamma$ leaves an elusive issue in understanding the novel topological
characteristics in LaBi.

The detailed photon energy dependent study summarized in Fig. 3 is performed from 20 to 100 eV to demonstrate the 2D nature of the SSs. We
show the FS mapping data in the $k_y$-$k_z$ plane at $E_F$ and $E$ = -0.14 eV, corresponding to the DP at $X$, with different photon energies
in Figs. 3(a) and 3(c), respectively. According to the photon energy dependence measurements, we estimate the inner potential of LaBi to 14
eV. The $k_z$ = 0 and $\pi$ planes at the BZ center correspond to $h\nu$ = 51 and 82 eV, respectively, while at the BZ boundary the values
are $h\nu$ = 53 and 84 eV, respectively. As illustrated in Fig. 3(a), the two selected photon energies ($h\nu$ = 53 and 84 eV) for the
investigations on the electronic structure at $k_z$ $\sim$ 0 and $\pi$ planes, respectively, are reasonable. We can clearly identify the
two SSs around the $\Gamma$ and $X$ points showing no obvious dispersion along the $k_z$ direction from the $k_y$-$k_z$ maps in Figs. 3(a)
and 3(c). We focus on the massless Dirac-like SS around $X$ and plot the MDCs at $E_F$ and $E$ = -0.14 eV in Figs. 3(b) and 3(d), respectively.
The $k_z$ independence of the Fermi crossings and DP provide further proof for the 2D surface band nature of this SS.

\begin{figure}[!t]
  \setlength{\abovecaptionskip}{-0.3cm}
  \begin{center}
  \includegraphics[width=0.85\columnwidth]{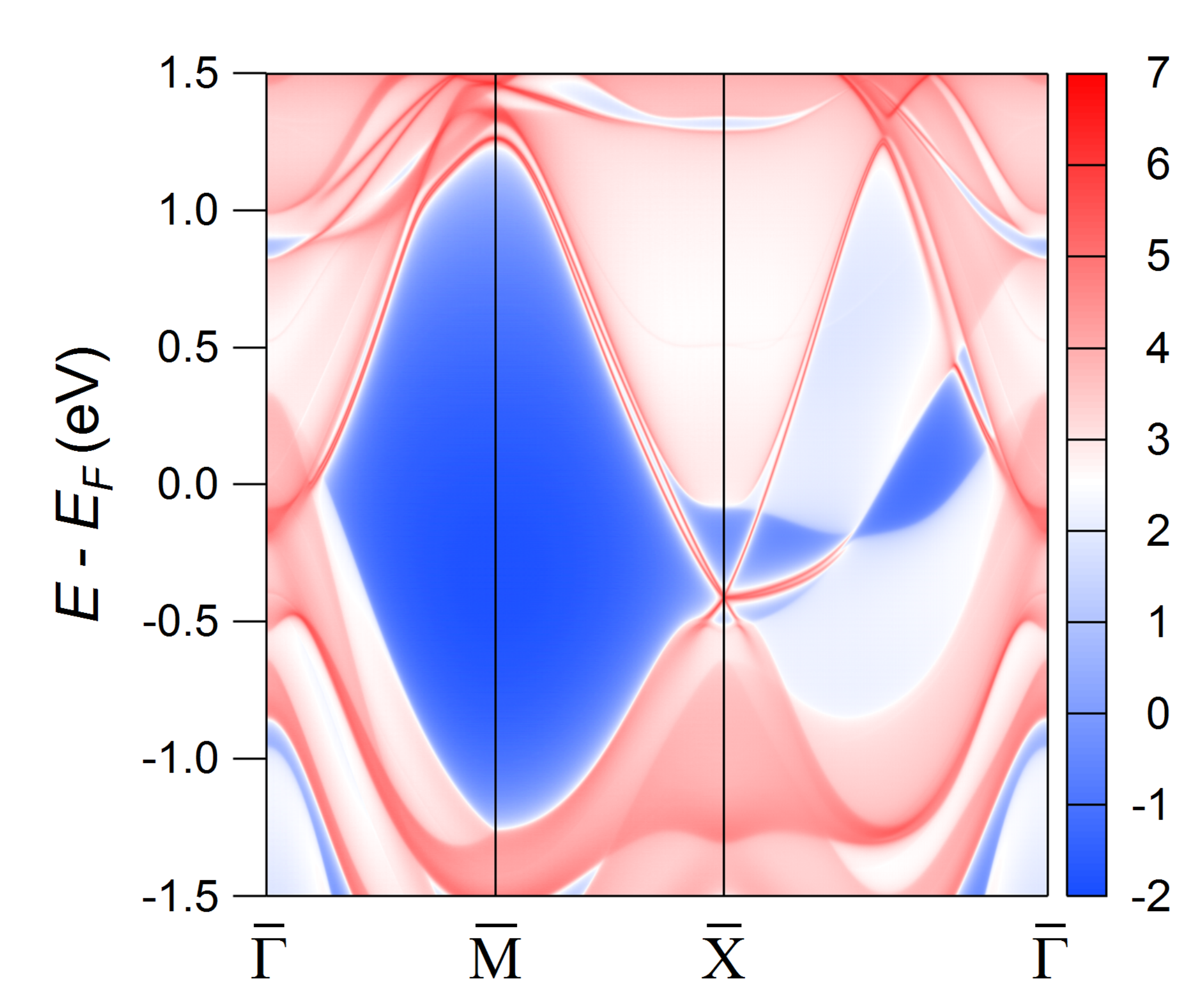}
  \end{center}
  \caption{(Color online)
  Calculated band structure along $\bar{\Gamma}$-$\bar{M}$-$\bar{X}$-$\bar{\Gamma}$ for a semi-infinite slab of LaBi. The sharp red
  curves represent the SSs for (001) surface, whereas the shaded regions show the spectral weight of projected bulk bands.
  }
\end{figure}

We also calculate the (001) surface band structure for a semi-infinite slab along the $z$ direction by using the Green's-function method, 
which is based on the TB Hamiltonian. As shown in Fig. 4, although one cannot unambiguously determine the SSs around $\bar{\Gamma}$ due 
to the convoluted bulk states, the Dirac-like SSs around $\bar{X}$ are consistent with our experimental results. The additional surface 
bands should relate to the topologically trivial origin.

We notice that there are two recent ARPES studies focusing on the surface band structure and topological properties of LaBi
\cite{Niu2016,Nayak2016}, both reporting the presence of two separated Dirac cones at $X$. Nevertheless, the good consistency
between our experimental band dispersions and slab model calculations demonstrate the existence of only one Dirac cone at $X$.
Further, we perform a similar semi-infinite slab calculation terminated by the surface on the other side of cleavage, and present
the result into the Supplemental Material \cite{Supplementary}. One can resolve two well-separated Dirac cones located at $\bar{X}$
from this surface termination. Although both sides of the cleavage should have identical atomic arrangement, the different calculated
surface band structures make this issue much more controversial \cite{NbAsHasan}. Therefore, we speculate that the distinct observations
at $X$ might result from the different surface terminations and periodic potentials after cleavage. We believe that further studies should
be invested on the topological characteristics of LaBi both theoretically and experimentally.

To conclude, we have performed ARPES experiments and first-principles calculations to investigate the electronic structure and intrinsic
topological characteristics of LaBi. A nontrivial band anticrossing is obviously observed along the $\Gamma$-$X$ direction. We further
identify two distinct topological SSs. One is gapless around $X$ and the other is massive around $\Gamma$ with a $\sim$23-meV gap. Such
an exotic mass acquisition of Dirac fermions requires future studies to clarify. Developed theoretical calculations need to be formulated
to resolve the surface band at $\Gamma$ from the convoluted bulk states. The presence of band inversion and odd number of massless Dirac
cones provide compelling evidence that LaBi is analogous to the time-reversal $Z_2$ TI. The rich and novel topological states in LaBi and
other compounds of the Ln$X$ family offer an excellent opportunity for investigating the evolution of topological properties from the
perspective of TPT, facilitating the future search of new topological phases.

This work was supported by the Ministry of Science and Technology of China (Programs No. 2012CB921701, No. 2013CB921700, No. 2015CB921000, No.
2016YFA0300300, No. 2016YFA0300504, No. 2016YFA0300600, No. 2016YFA0302400, and No. 2016YFA0401000), the National Natural Science Foundation of
China (Grants No. 11274381, No. 11274362, No. 11474340, No. 11234014, No. 11274367, No. 11474330, No. 11574394, and No. 11674371), and the Chinese
Academy of Sciences (CAS) (Project No. XDB07000000). RL and KL were supported by the Fundamental Research Funds for the Central Universities, and
the Research Funds of Renmin University of China (RUC) (Grants No. 17XNH055, No. 14XNLQ03, No. 15XNLF06, and No. 15XNLQ07). Computational resources
have been provided by the Physical Laboratory of High Performance Computing at RUC. The FSs were prepared with the XCRYSDEN program \cite{Kokalj2003}.
YH was supported by the CAS Pioneer Hundred Talents Program.

\end{document}